\documentclass[twocolumn]{article}

\usepackage{arxiv}

\usepackage[utf8]{inputenc} 
\usepackage[T1]{fontenc}    
\usepackage[hidelinks]{hyperref}       
\usepackage{url}            
\usepackage{booktabs}       
\usepackage{amsfonts}       
\usepackage{nicefrac}       
\usepackage{microtype}      

\usepackage{xcolor}
\definecolor{darkgreen}{rgb}{0.0, 0.5, 0.0}
\definecolor{amaranth}{rgb}{0.9, 0.17, 0.31}
\definecolor{azure}{rgb}{0.0, 0.5, 1.0}

\usepackage{listings}

\usepackage{graphicx}
\usepackage{caption}
\usepackage{subcaption}

\usepackage{xspace}

\usepackage{siunitx}

\usepackage{paralist}

\usepackage{aas_macros}

\newcommand{\orcid}[1]{\href{https://orcid.org/#1}{\includegraphics[height=10pt]{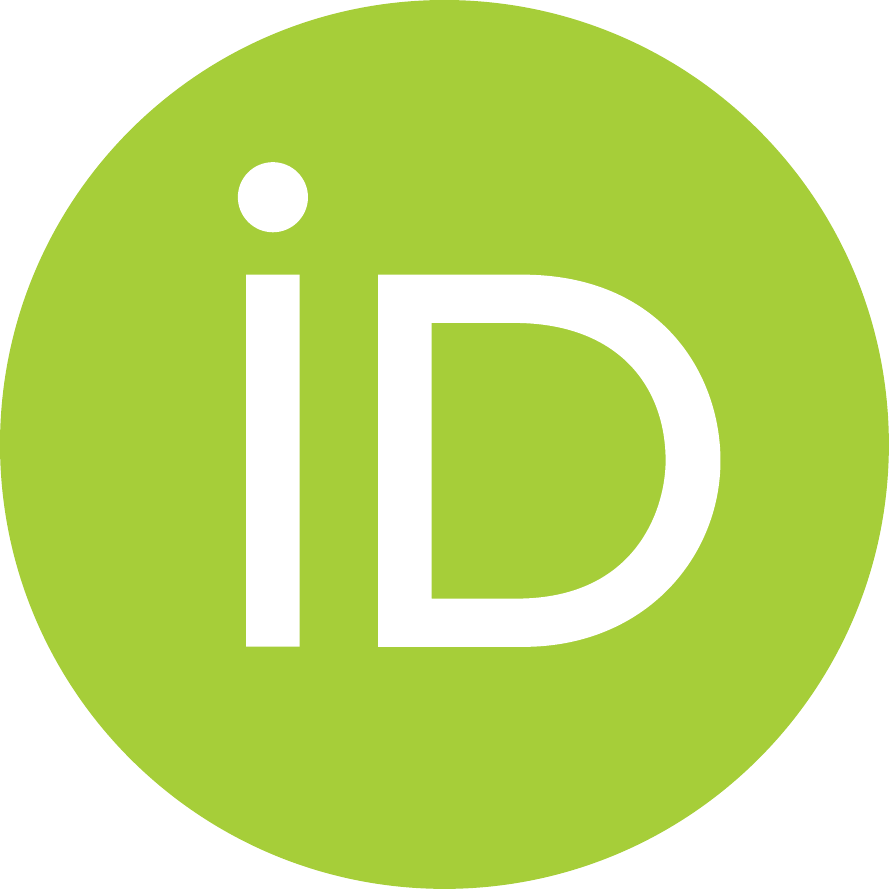}}}
\newcommand{\octo}{Octo-Tiger\xspace}
\newcommand{\octotiger}{Octo-Tiger\xspace}
\newcommand{\amt}{Asynchronous Many Task\xspace}

\newcommand{\futurization}{Futurization\xspace}

\title{Performance Measurements within Asynchronous Task-based Runtime Systems: \\ \Large A Double White Dwarf Merger as an Application}

\author{
  Patrick Diehl\orcid{0000-0003-3922-8419}, Dominic Marcello, Parsa Amini, Hartmut Kaiser\orcid{0000-0002-8712-2806} \\
  Center of Computation and Technology\\
  Louisiana State University\\
  Baton Rouge, LA, U.S.A. \\
  \texttt{\{pdiehl,hkaiser,parsa\}@cct.lsu.edu $|$ dmarrce504@gmail.com} \\
   \And
 Sagiv Shiber\orcid{0000-0001-6107-0887}, Geoffrey C. Clayton\orcid{0000-0002-0141-7436}, Juhan Frank \\
  Department of Physics \& Astronomy\\
   Louisiana State University\\
 Baton Rouge, LA, U.S.A. \\
  \texttt{frank@phys.lsu.edu} $\vert$ sshiber1@lsu.edu  $\vert$ gclayton@fenway.phys.lsu.edu \\
  \AND
   Gregor Dai\ss, Dirk Pfl\"uger\orcid{0000-0002-4360-0212} \\
   \textit{IPVS} \\
  University of Stuttgart, Stuttgart, Germany \\
   \texttt{\{Dirk.Pflueger,Gregor.Daiss\}@ipvs.uni-stuttgart.de} \\
   \And
   David Eder, Alice Koniges \\
   Maui High Performance Computing Center \\
   University of Hawaii, Maui, HI, U.S.A \\
   \texttt{\{dceder,koniges\}@hawaii.edu} \\
   \And
   Kevin Huck\orcid{0000-0001-7064-8417} \\
   \textit{OACISS} \\
   University of Oregon, Eugene, OR, U.S.A. \\
   \texttt{khuck@cs.uoregon.edu} \\
}

\begin{document}

\onecolumn

\maketitle

\begin{abstract}
Analyzing performance within asynchronous many-task-based runtime systems is challenging because millions of tasks are launched concurrently. Especially for long-term runs the amount of data collected becomes overwhelming. We study HPX and its performance-counter framework and Autonomic Performance Environment for Exascale to collect performance data and energy consumption. We added HPX application-specific performance counters to the \octo full 3D adaptive multigrid code astrophysics application. This enables the combined visualization of physical and performance data to highlight bottlenecks with respect to different solvers. We examine the overhead introduced by these measurements, which is around 1\%, with respect to the overall application runtime. We perform a resolution study for four different levels of refinement and analyze the application's performance with respect to adaptive grid refinement. The measurements' overheads are small, enabling the combined use of performance data and physical properties with the goal of improving the code's performance. All runs were obtained on NERSC's Cori, Louisiana Optical Network Infrastructure's QueenBee2, and Indiana University's Big Red 3.
\end{abstract}


\twocolumn

\section{Introduction}
Over the last decade, the development of asynchronous task-based runtime systems has offered users an alternative to MPI. These systems provide them abstractions to distribute their applications on large systems.
However, while these runtime systems can handle millions of concurrent tasks over thousands of compute nodes, collecting information about application performance and framework overhead is a challenge, in particular for massively parallel large-scale simulations.

In this work, we demonstrate how we can gather performance data for massively parallel production runs on NERSC's Cori, Louisiana Optical Network Infrastructure's QueenBee\num{2}, and Indiana University's Big Red \num{3}.

We use the asynchronous task-based runtime system HPX and its performance-counter framework, and the extension Automatic Performance for Exascale (APEX) to collect the data. Furthermore, we discuss the overheads encountered and physics results obtained for a double white dwarf merger simulation.

As an application, we choose \octo.
\octo is a code for the study of self-gravitating astrophyscial fluids using Newtonian gravity. It is particularly suited for stellar mergers, and we use it to simulate a merger in this paper.
\octo is based entirely upon HPX, which provides the performance counters to profile our application in a massively parallel setting.
Of the many types of stellar interactions, the merger event is the most compelling. The merging process is violent, and can usually be seen in many wavelength regimes as a bright outburst, observable at large distances. Some classes of stars and outbursts, for example type Ia supernovae, derive almost certainly from mergers; for others, such as magnetic white dwarfs, mergers are suspected. None of these phenomena is well understood. For the simulations in this paper, we investigate a stellar merger that may form an R Coronae Borealis (RCB) star.

Previously, it has been shown that \octo can scale up to thousands of compute nodes~\cite{daiss2019piz}.
Further, the performance of individual kernels has been showcased~\cite{daiss2019piz}, and \octo has been used to measure the overhead of certain HPX components before~\cite{8956479}. 
\octo is already used for production runs, making it an ideal candidate for performance and overhead measurements in a real-world scenario.

Here, we demonstrate how to collect and visualize both physics and performance data, including energy consumption. Allowing us to identify bottlenecks and to optimize on the level of algorithms and on a lower code level.

 In this paper we perform a resolution study using a full-scale white dwarf merger runs with \octo. This gives us the insight for the integration of additional physical properties and refined simulations for better production runs.

From the computational perspective, we collect a multitude of performance measurements about APEX, HPX, and \octo itself during the full-scale resolution study and several complementary short runs. This enables us to gain insight into the overheads of the utilized frameworks. 
These measurements include the overhead of taking the APEX measurements, the amount of energy required for the simulation at different scales, the HPX idle rate, and the HPX AGAS (Active Global Address Space).
Some of those measurements we collect during short runs since \octo must be run with multiple configurations to gather the data. 
Other measurements, like the ones for the AGAS, are collected during the actual resolution study. While there had been similar measurements of the AGAS in Amini et. al.~\cite{8956479}, these were only for short simulation runs.
We extend their results and obtain measurements for long, real-world production runs as well. In contrast to the short runs, long runs exhibit highly dynamic behavior and require frequent adaptive mesh refinement, which poses extra challenges for scalability and efficient parallelization.

Here we present both new simulation results of \octo and the overhead measurements of the utilized frameworks.

\section{Related Work}
\label{sec:work}

\subsection{R Coronae Borealis (RCB) merger }
When two stars merge into one, they form objects with strange characteristics.
The R Coronae Borealis are rare, carbon stars with little hydrogen and dust-induced variability~\cite{Clayton2012}. They are almost impossible to explain through single-star evolution. The observational evidence points to a stellar merger of two white dwarfs (WDs). The RCB stars are close cousins of Type Ia SNe.

WD-WD modeling is dominated by the question of where and when the primary star detonates. But the mergers that produce RCB stars are low mass and the star does not explode. Previous simulations have been important to studies of RCB stars~\cite{2019MNRAS.488..438L}.

\subsection{Performance measurements and their visualization}
Typically, performance data is visualized and represented in the physical
and/or logical context of the hardware and/or software resources used in the
simulation. Data is organized by processes and threads, and visualized with
respect to nodes, network topology and CPU architectures. Huck et al.~\cite{Huck2014vpa} integrated performance data with simulation output
in order to project the performance data into the scientific domain.
Using the Scalable Observation System (SOS),
performance data was aggregated and
queries were executed to extract performance data and
generate Visualization Toolkit (VTK) output files~\cite{wood2019}. Using a similar approach, fusion
simulation performance data was aggregated and exported to VTK
files~\cite{choi2018}.

\section{Scientific application: unstable mass transfer in a double white dwarfs system leading to a merger}
\label{sec:application}
Approximately half of the double white dwarf systems in our galaxy will merge within a Hubble time.
These systems slowly lose their orbital energy by emitting gravitational waves. This drives the two stars closer together, until they start to interact via mass transfer and ultimately merge. Observational evidence suggests the stellar merger scenario for the origin of RCB stars~\cite{Clayton2012}. We simulate the interaction between two low mass white dwarfs that leads to merger.

\subsection{Transfer of angular momentum}
\label{subsec:angularMomentum}
Mass transfer in systems with mass ratios above a certain threshold results in merger due to tidal disruption of the donor star. As angular momentum is transferred from the orbit to the spin of the accretor, the orbital angular momentum decreases. For certain equations of state, the donor increases in size as it loses mass, leading to runaway mass transfer. The donor star is ripped apart by tidal forces, its remnants falling onto the accretor. 

According to previous simulations, the onset of mass transfer and the time the binary system experiences mass transfer before merging are highly sensitive to initial conditions. Motl et. al.~\cite{Motl_2017} showed that the resolution of a simulation affects the initial mass transfer, which in turn determines the length of the mass transfer episode before the merger. The higher the resolution, the longer the duration before a merger. We expect to see the same behavior in our simulations. 

The consistent simulation of mass transfer and its feedback on the binary dynamics is a serious challenge for every numerical method. For smoothed particle hydrodynamics (SPH) codes, this is because the transferred matter comes initially from the WD surface which is the poorest resolved region of the star. For finite volume codes, this is due to the resolution-dependent angular momentum conservation. \octo is a good candidate in overcoming these challenges because it is a finite-volume code that conserves angular momentum to high precision.

\subsection{The stellar model}
\label{stellarModel}
Our initial model is generated using \octo's Self Consistent Field (SCF) code module. The initial masses of the accretor and donor are $0.5 M_\odot$ and $0.35 M_\odot$ (mass ratio of $0.7$), respectively, with
an initial separation of $0.047 R_\odot$ and an initial period of $111 \mathrm{s}$. The accretor is composed of a $0.455 M_\odot$ core with an even mixture by mass of carbon and oxygen (CO), surrounded by a helium shell of mass $0.045 M_{\odot}$. This helium shell has a role in reducing the presence of $^{16}$O in the outer layers, where the nuclear reactions occur. The donor is composed of only helium.

The grid domain runs from $-6.12 R_\odot$ to $+6.12 R_\odot$ in each dimension. Table~\ref{tab:levels} shows
the finest grid cell size of each maximum level of refinement.
We approximate the stars as $n=3/2$ polytropes and evolve the gas with an ideal gas equation of state. 

\begin{table*}[tb]
 \caption{The four levels of refinements for resolutions utilized. }
 \label{tab:levels}
 \centering
 \begin{tabular}{cllllll|ll}
 \toprule
 Levels of & Initial & Finest grid & Finest Grid & Initial & AMR & Memory & Time of & Diff from \\
 Refinement & Sub-grids & Cell Size ($\mathrm{cm}$) & Cell Size ($R_\odot$) & Leaf Nodes & Boundaries & (GB) & Disruption & Next Level \\\midrule
 \num{10} & \num{1185} & $1.03 \times 10^8$ & $1.48\times 10^{-3}$ &\num{1038} &\num{1503} & \num{15.6} & 6.7 & 6.7 \\
 \num{11} & \num{4974} & $5.18 \times 10^7$ & $7.40 \times 10^{-3}$ &\num{4352} &\num{4158} & \num{66.6} & 13.4 & 2.1 \\
 \num{12} & \num{8781} & $2.59 \times 10^7$ & $3.70 \times 10^{-4}$ &\num{5969} & \num{7683} & \num{117.7} & 15.5 & 1.3 \\
 \num{13} & \num{25353} & $1.26 \times 10^7$ & $1.85 \times 10^{-4}$ & \num{22184} & \num{11800} & \num{339.8} & 16.8 & -- \\\bottomrule
 \end{tabular}
\end{table*}

\section{Software framework}
\label{sec:framework}

\subsection{\octo}

\octotiger is a finite volume, octree based, adaptive mesh refinement (AMR) code that solves the in-compressible Euler equations with self-gravity on a fully three-dimensional rotating Cartesian mesh. It uses a 3rd order method to solve the hydrodynamics and a fast multipole method (FMM) to solve the gravity. The task-based approach provided by HPX is ideally suited for an application such as Octo-Tiger for several reasons. Through futures, HPX provides the ability to set up tasks dependent on data availability. HPX schedules those tasks as the data become available. The global address space of HPX allows Octo-Tiger to easily set up its octree over the computational domain and allows for easy load balancing. Finally, HPX futures do not distinguish between node level and system level parallelism, removing the need to use OpenMP to parallize at the node level. We refer the reader to Marcello, Shiber, et. al. ~\cite{marcello2021octo} for a more detailed description of \octo.

\subsection{HPX}
\label{subsec:hpx}

The development of \octotiger in ISO C++11 using HPX~\cite{Kaiser2020} is shown in by Dai\ss et. al. ~\cite{daiss2019piz}.
HPX is a C++ standard library for distributed and parallel programming built on
 top of an \amt (AMT) runtime system. Such AMT runtimes may provide a means
 for helping programming models to fully exploit available parallelism on
 emerging HPC architectures. The HPX methodology described here includes
 the following components:

\begin{compactitem}
\item An ISO C++ standard conforming API that enables wait-free asynchronous parallel programming, including futures, channels, and other primitives for asynchronous execution.
\item An Active Global Address Space (AGAS) that supports load balancing via object migration and enables exposing a uniform API for local and remote execution.
\item An active-message networking layer that enables running functions close to the objects they operate on. This also implicitly overlaps computation and communication.
\item A work-stealing lightweight task scheduler that enables finer-grained parallelization and synchronization and automatic load balancing across local compute resources.
\item APEX, an in-situ profiling and adaptive tuning framework.
\end{compactitem}
\smallskip

HPX exposes an asynchronous, standards conforming programming model enabling \futurization,
with which developers can express dataflow execution trees that generate billions of HPX tasks that are scheduled to execute only when their dependencies are satisfied. Also, \futurization\ enables automatic parallelization and load-balancing to emerge. This provides a unified approach to intra- and inter-node parallelism based on proven generic algorithms and data structures available in today's ISO C++ Standard. The programming model is intuitive and enables performance portability across a broad spectrum of diverse HPC hardware.

Additionally, HPX provides a performance counter and adaptive tuning framework that allows users to access performance data, such as core utilization, task overheads, and network throughput; these diagnostic tools were instrumental in scaling \octo to the full machine.

\subsubsection{AGAS}
AGAS, short for Active Global Address Space, is the part of HPX runtime system
that provides the means to access objects that live on different compute nodes
from other compute nodes.
Through this system, accessing a global object is transparent to the code and
is done through the same API, whether the object lives on the same node as the
code that attempts to access it or not.
AGAS is able to hide some I/O and object access latencies through the use of
active messages when accessing remote objects.

Since AGAS is an active component of the HPX runtime system and does its work during
application execution, it takes a portion of the execution time.
While AGAS overheads have been shown to be insignificant in short tests on Piz Daint at CSCS~\cite{8956479},
this study provides an opportunity to study its overheads in a long-running
application with dynamic adaptivity and observe whether it affects application execution times in an
interesting fashion.

\subsubsection{Performance counters}

\begin{lstlisting}[
 basicstyle=\small\ttfamily,
 caption=Example for an application-specific HPX performance counter and its registration,
 label={lst:hpxappcounter},
 float=tb,
 numbers=left,
 escapechar=|,
 tabsize=2,
 numberstyle=\tiny,
 numbersep=5pt,
 belowskip=-0.8 \baselineskip
 ]
// Function to return the application-specific
// performance counter
uint64_t leaf_count(bool reset) {|\label{line:func}|
 std::lock_guard<hpx::mutex> lock(leaf_count_mtx);
 if (reset) {
 cumulative_leaf_count = 0;
 }
 return cumulative_leaf_count;
}
// Register the performance counter
void register_counter() {
 using namespace hpx::performance_counters;
 install_counter_type(|\label{line:reg}|
 "octotiger/subgrid_leaves",|\label{line:name}|
 &leaf_count,
 "total number of subgrid leaves processed");
}
 \end{lstlisting}

HPX performance counters are first class objects, each with a global address mapped to a unique symbolic name, useful for introspection at execution time by the application or the runtime system.
The performance counters are used to provide information about how the runtime system or the application is
performing. Counter data can help determine system bottlenecks and fine-tune system and application performance.

HPX exposes special API functions that allow one to create, manage, and read the counter data~\cite{conf/ipps/GrubelKHC16}. Any performance counters can be accessed remotely or locally using the same API. Moreover, since all counters are accessible through a uniform set of functions, any code consuming counter data can be utilized to access arbitrary system information with minimal effort. HPX also allows the extension of the set of performance counters to include application-specific information (see Listing~\ref{lst:hpxappcounter} for an example of all code necessary to do so). An application needs to expose the counter data through a function (line~\ref{line:func}) that has to be registered with HPX (line~\ref{line:reg}) with a unique name (line~\ref{line:name}). The application-specific performance counters are then exposed through the same interfaces as the predefined counters and are readily available to APEX and other tools for intrinsic performance analysis.

\subsection{APEX}
APEX~\cite{huck2015autonomic} (Autonomic Performance Environment for Exascale)
is a performance measurement library for distributed, asynchronous multitasking
systems such as HPX. It provides lightweight measurement and high concurrency. To support performance measurement in systems that employ user level threading, APEX uses a dependency chain rather than the call stack to produce traces. APEX supports both synchronous and asynchronous introspection. The synchronous measurement by APEX uses a timer-based instrumentation and event listeners. The API includes events to start, stop, yield, or resume
timers and to sample values in the application. In contrast, the asynchronous measurements periodically perform first- and third-person observations of the application, libraries, the hardware and operating system states such as load,
energy consumption and resource utilization.

The \textit{policy engine} within APEX provides an API to construct policies
that can modify the behavior of the application, execute a function in
the runtime or select important runtime and application parameters. Typically,
the policies are used to apply optimization strategies but can be implemented as
a feedback-and-control system. There are two ways to execute a policy: either
explicitly triggered or asynchronously periodic. A triggered policy can be
initiated by a specific event within the HPX runtime. In contrast, periodic policies use a periodic timer interrupt which is specified during the policy's registration. All policies are stored in a policy
queue and executed in the order they are registered. Defined policies can search for a set of optimum parameters
by minimizing a measurement value from APEX, such as the time of a measured
region/task or by looking at any other introspection data gathered by APEX.

\subsection{PAPI}
Both APEX and HPX are integrated with PAPI (The Performance
API). PAPI provides a unified, generic interface to CPU hardware performance
counters across many hardware platforms and operating systems. PAPI provides
portable, abstracted counter names to access the hardware counters without
having to customize the measurement to every system available. Recently, PAPI
has also added components to measure GPU hardware as well as overall
system measurements such as network interfaces, file system performance,
sensor data, and energy consumption.

\section{Computer Science Results}
\label{sec:results}
This section contains the results of the computer science and the domain science aspects for the RCB merger described in Section~\ref{sec:application}. Table~\ref{tab:levels} lists the details of the four different level of refinements of the RCB scenario used for the resolution study and the investigation of performance.

\subsection{Overhead measurements}
This section studies the overhead of HPX's performance counters and APEX/PAPI. On NERSC's Cori system the sub-grids processed per second for the first two levels were obtained up to \num{1024} nodes. For one node up to \num{64} nodes four different scenarios were executed: \textit{I} HPX + performance counters and APEX, \textit{II} HPX + APEX, \textit{III} HPX + Performance counters, and \textit{IV} pure HPX.
For up to \num{1024} nodes and all long-term runs the configuration \textit{I} was used for producing the results in the following sections.

For all four configurations, the amount of processed sub-grids per second only differs by approximately \num{1}\si{\percent}, which is acceptably within the expected variability of a typical shared system.
In general adding HPX's performance counters and APEX does not affect the sub-grids processed per second and no huge effect on performance can be identified up to \num{64} nodes. Note that a run using the performance counters and APEX/PAPI is only needed once to identify potential bottle necks in the application by combining HPX specific and application-specific performance counters. We had to choose four small enough levels fitting into our awarded node hours for the resolution study presented in this paper.
The scaling flattens out soon, since we had not enough work for the larger amount of nodes.

\subsection{Performance counters}

\subsubsection{HPX counters}
\begin{figure}[tb]
	\centering
	\includegraphics[width=0.85\linewidth]{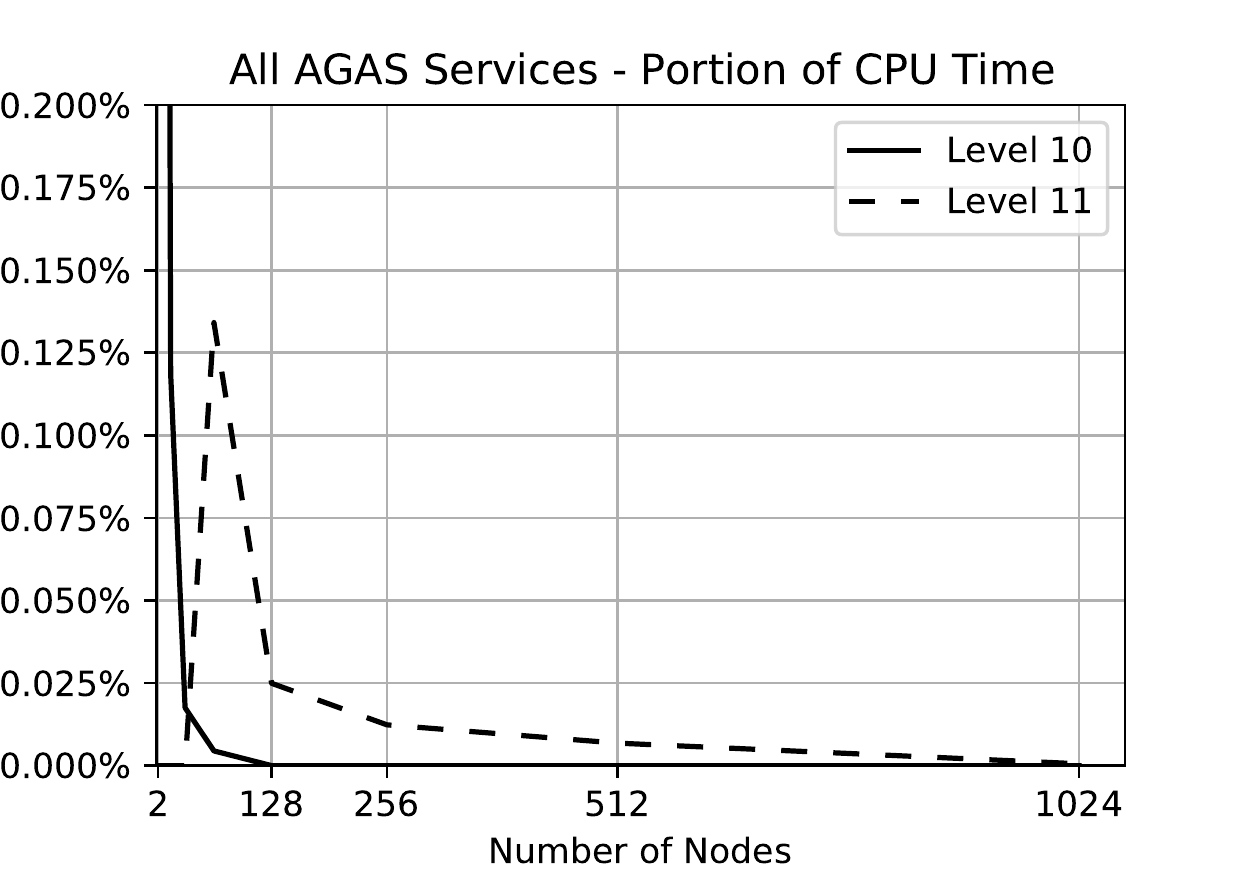}
	\caption{Portion of CPU time spent executing AGAS code. These runs were done on NERSC's Cori.}
	\label{fig:agas-scaling-exec-times}
\end{figure}

Fig~\ref{fig:agas-scaling-exec-times} shows that the percentage of CPU time
that is spent executing AGAS code is
insignificant. This is the same behavior has been previously reported~\cite{8956479}
for a different scenario on Piz Daint. In this study we conduct a broader study
of AGAS overheads by measuring them on four long-term production runs from \num{45}, \num{177.50}, \num{227.83} hours up to \num{14} days on QueenBee2. The AGAS overhead (\% in parenthesis) was (\num{0.019}) \num{10} , (\num{0.12}) \num{20}, (\num{ 0.17}) \num{40}, and (\num{0.17}) \num{64}-\num{128} nodes. As with the short-term runs, we see that the
AGAS overheads are insignificant.

\subsubsection{APEX counters}
\begin{figure}[tb]
	\begin{subfigure}[b]{0.5\textwidth}
 		\includegraphics[width=0.85\linewidth]{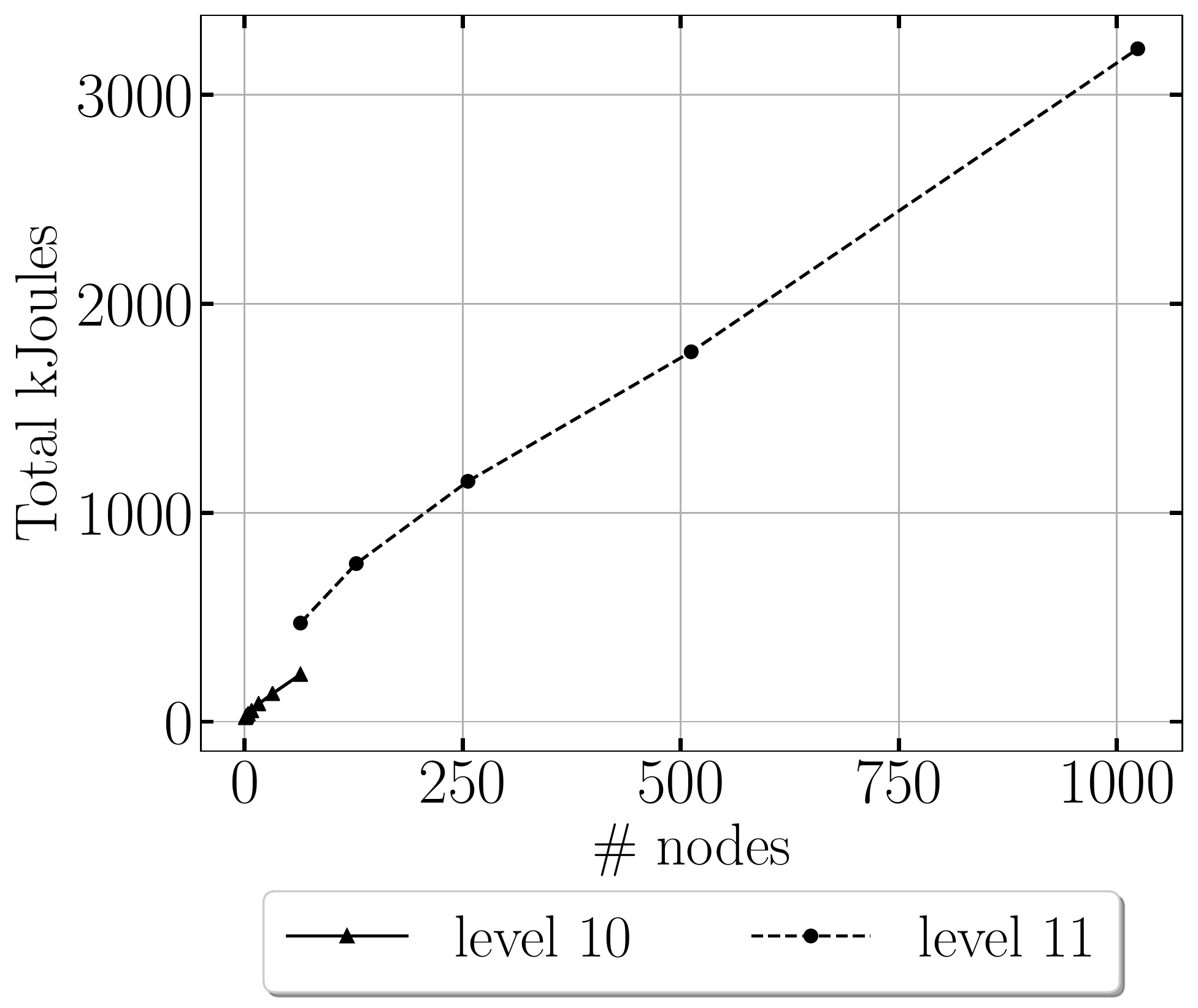}
 		\caption{}
 		\label{fig:cori-scaling-kjoules}
	\end{subfigure}
	\hfill
	\begin{subfigure}[b]{0.5\textwidth}
 		\includegraphics[width=0.85\linewidth]{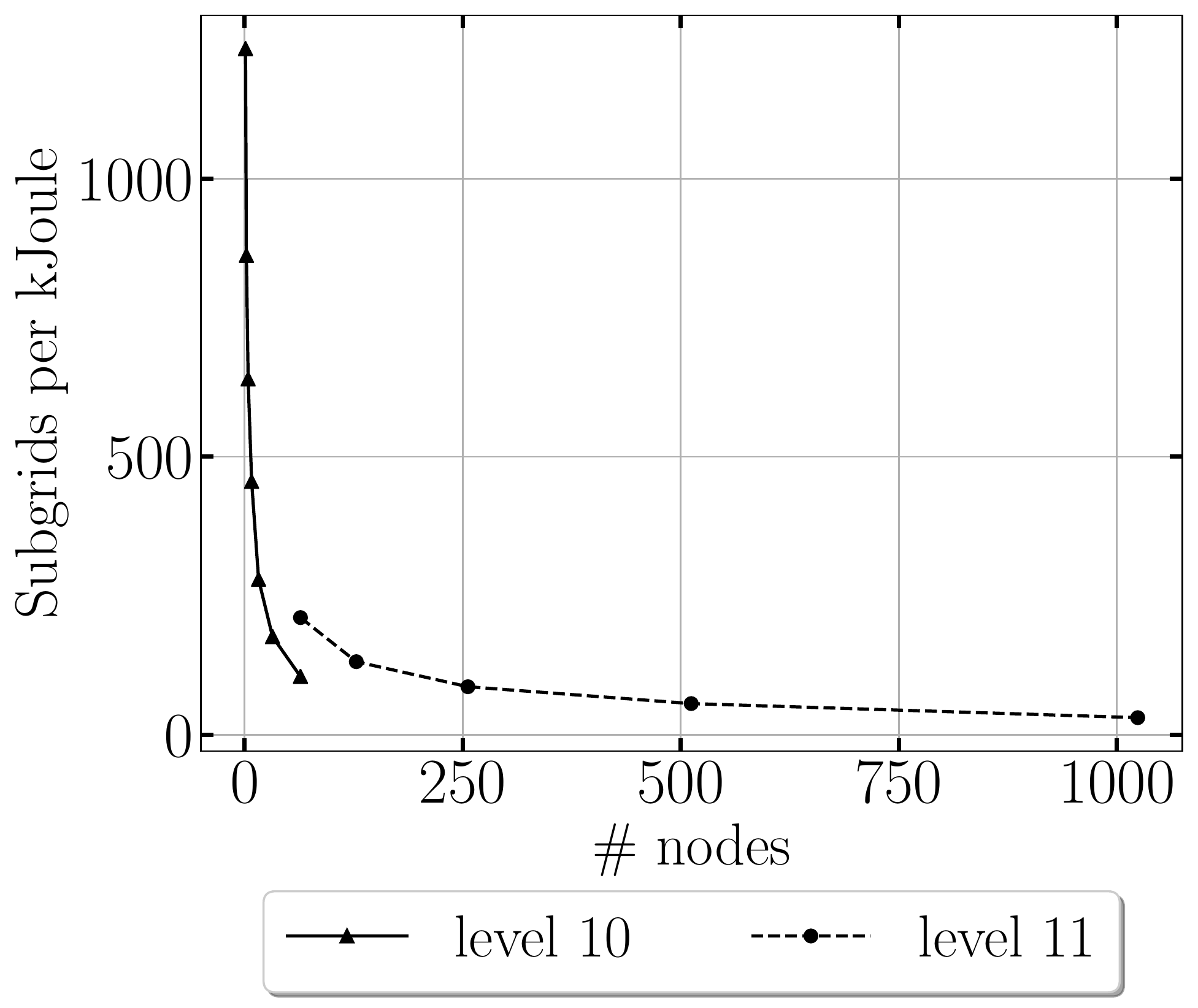}
 		\caption{}
 		\label{fig:cori-scaling-subgrids-per-kjoule}
	\end{subfigure}
 \caption{Total kJoules consumed for each run on Cori, and the number
 of sub-grids processed per kJoule for each run on Cori.}
	\label{fig:cori-energy-consumption}
\end{figure}

Our prospective metric to determine the optimal number of nodes is the
sub-grids processed per second. However, another metric could be to optimize
the energy consumption. Figure~\ref{fig:cori-scaling-kjoules} shows the total
energy consumption as the number of nodes increases and
Figure~\ref{fig:cori-scaling-subgrids-per-kjoule} shows the processed sub-grids
per \si{\kilo\joule} for a run on Cori, including initialization and shutdown. We could not perform the energy
measurements for the long-term runs, since the Linux kernel on QueenBee2 was too
outdated and this feature requires a kernel version $\ge 3.7$.

\subsubsection{\octo counters}
\label{sec:results:appcitaiton:counter}

	\begin{figure}[tb]
		\begin{subfigure}[b]{0.5\textwidth}
 		\includegraphics[angle=270,width=0.85\linewidth]{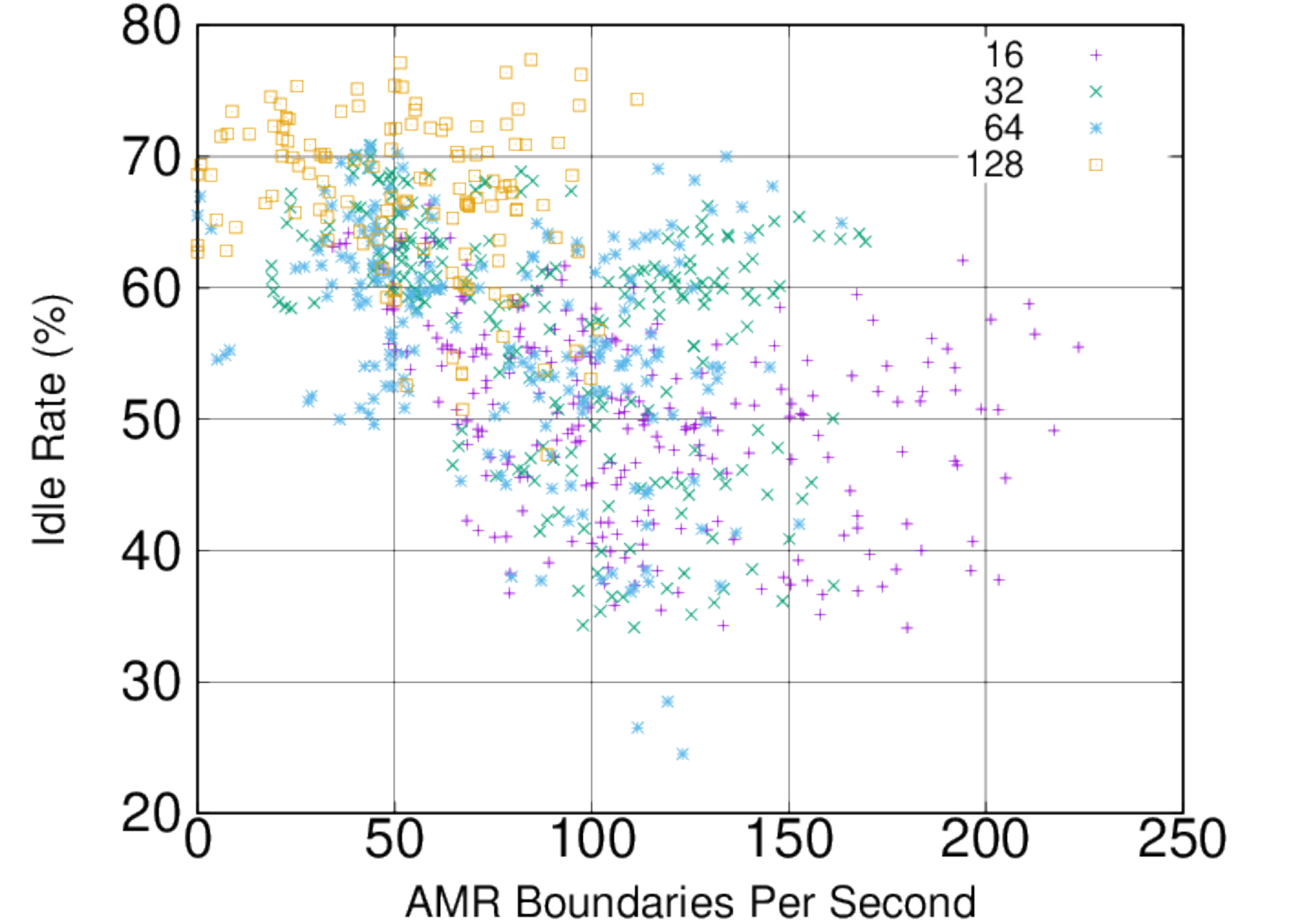}
 		\caption{}
 		\label{fig:scatter}
		\end{subfigure}
		\hfill
		\begin{subfigure}[b]{0.5\textwidth}
 		\includegraphics[width=0.85\linewidth]{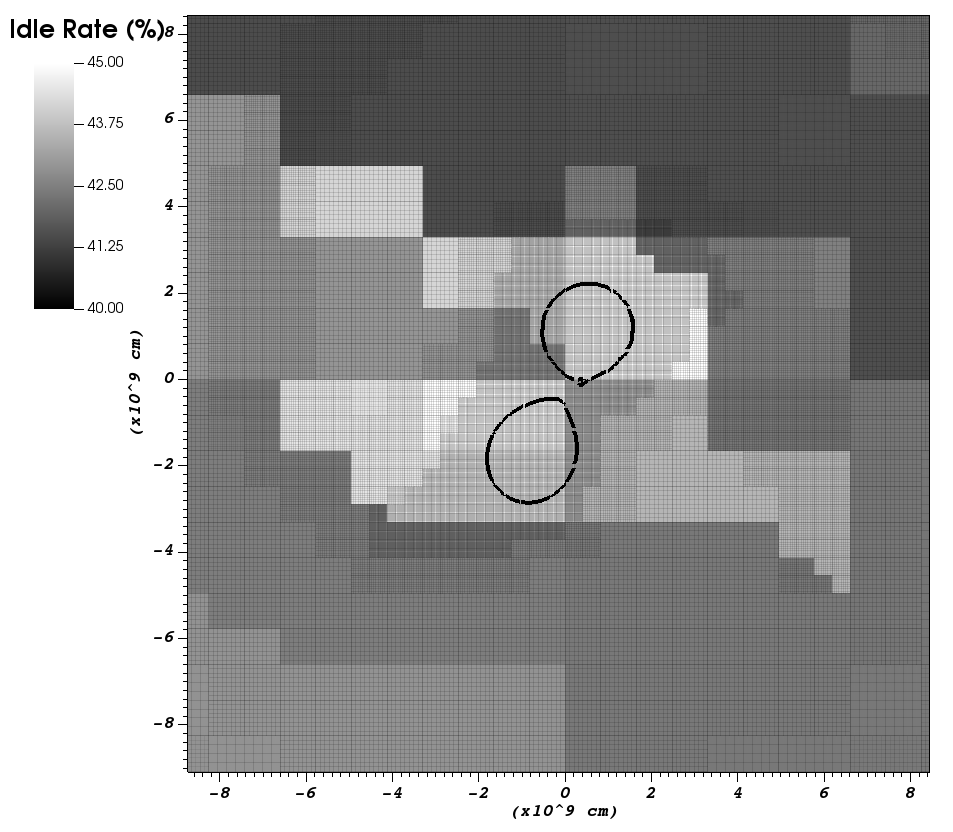}
 		\caption{}
 		\label{fig:idle_rates}
		\end{subfigure}
		\caption{Idle rate (\%) vs. AMR boundaries / s, sampled for each process every hour during part of the level 12 run (\subref{fig:scatter}) and the idle rates projected onto the computational domain (\subref{fig:idle_rates}). Higher idle rate indicates worker starvation due to an insufficient amount of work.}
	\end{figure}

\octo has three application-specific performance counters. They count the total numbers of sub-grids, leaf nodes, and AMR boundaries processed. HPX tracks the idle rate, defined as the
amount of time an HPX locality has no work ready to schedule, through its own performance counter. We show in Figure~\ref{fig:scatter} the idle rate versus the AMR boundaries processed per second for a short test run of the level \num{13} problem on Big Red 3 super computer. We ran the test on \num{16}, \num{32}, \num{64}, and \num{128} nodes. As expected, the runs with higher node counts have a higher idle rate since the amount of sub-grids for each node decays from \num{3169} to \num{198}. Despite the high idle rates, significant performance improvement in terms of total sub-grids per second is gained even moving from \num{64} (\num{3493}) to \num{128} (\num{4834}) nodes. Going from \num{16} to \num{128} nodes (\num{8}$\times$) gives a speedup of four for the processed sub-grids for a small problem size. We had to find a trade-off for running four levels of refinement for scientific results and could not run the optimal problem size with respect to scaling/permanence in this paper.

We also notice a weak dependence between the AMR boundary boundaries per second and the idle rate. Although the number of sub-grids and leaf nodes is distributed evenly between localities, the number of
AMR boundaries a given compute node must process varies between localities. Localities with more boundaries to process have lower idle rates due to their higher work load. These results suggest there may be an opportunity for more efficient load balancing by distributing the load based on AMR boundaries as well as sub-grid counts. In the future, another quantity to investigate with performance counters could be the ratio of local to remote AMR boundaries.

\octo also includes the idle rates by locality as part of its grid output. This feature can be used to investigate whether different parts of the spatial domain have lower or higher work loads. Figure~\ref{fig:idle_rates} shows the idle rates for a zoomed in equatorial slice of the domain for the level 12 run. There is a single density contour, and the AMR grid structure is shown. Since the idle rate is the same for each locality, the plot also reveals locality boundaries. The range of idle rates shown is very narrow, indicating good load balancing for this particular slice.


\section{Astrophysical Results}
\label{sec:result:convergence}
\begin{figure*}[tb]
	\begin{subfigure}[b]{0.3\textwidth}
 		\includegraphics[angle=270,width=\linewidth]{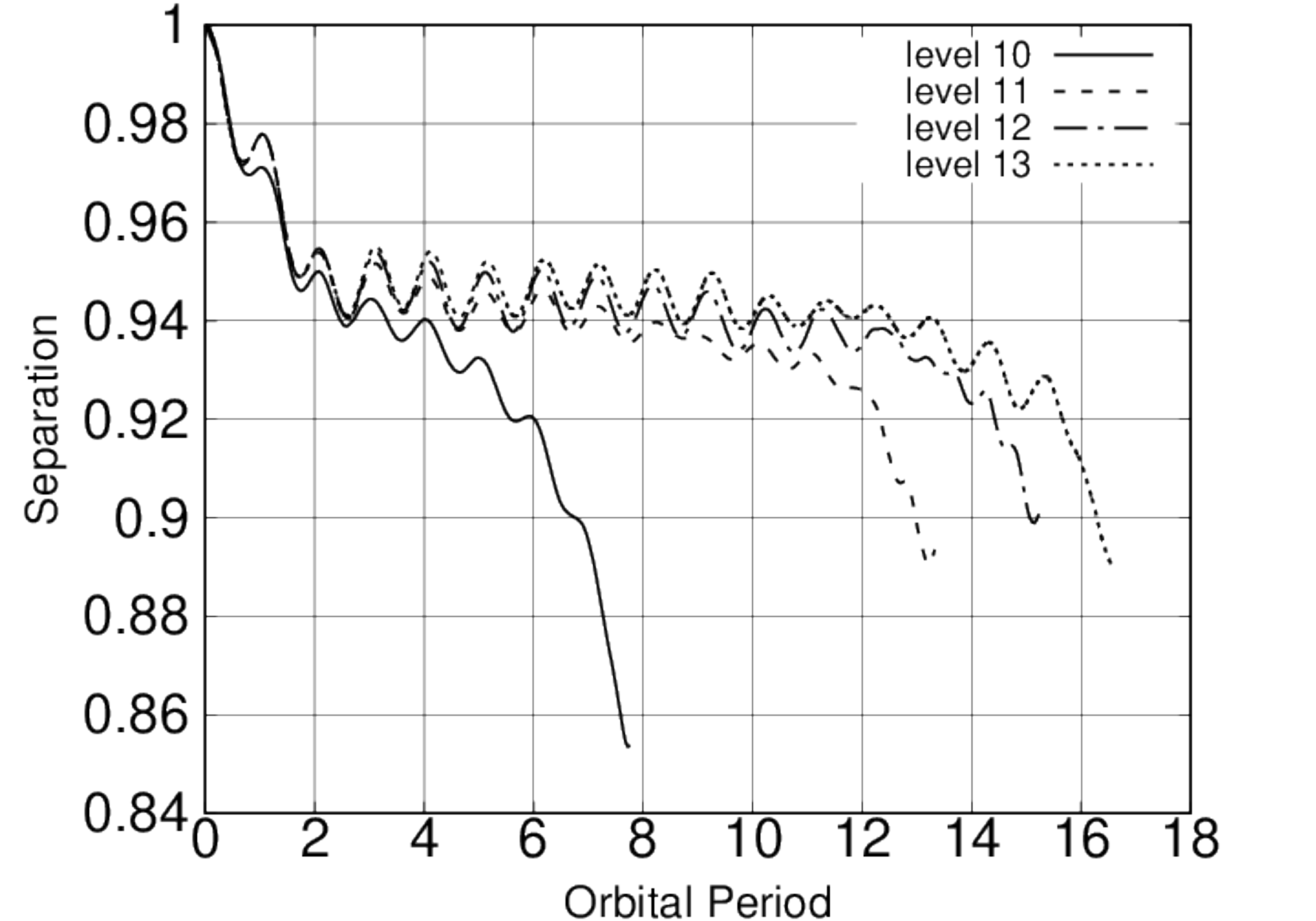}
		\caption{}
	\end{subfigure}
	\hfill
	\begin{subfigure}[b]{0.3\textwidth}
 		\includegraphics[angle=270,width=\linewidth]{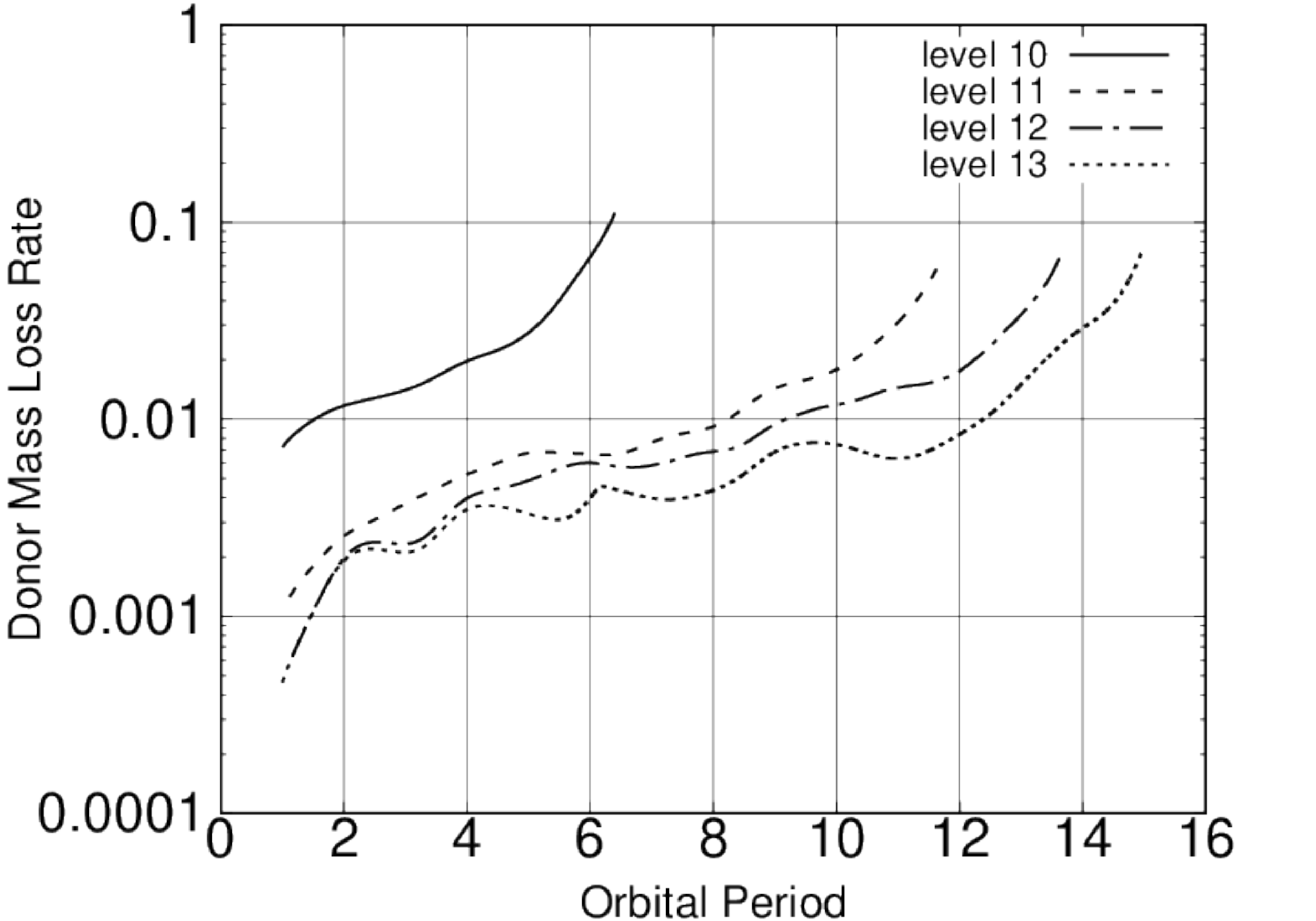}
		\caption{}
	\end{subfigure}
	\hfill
	\begin{subfigure}[b]{0.3\textwidth}
 		\includegraphics[angle=270,width=\linewidth]{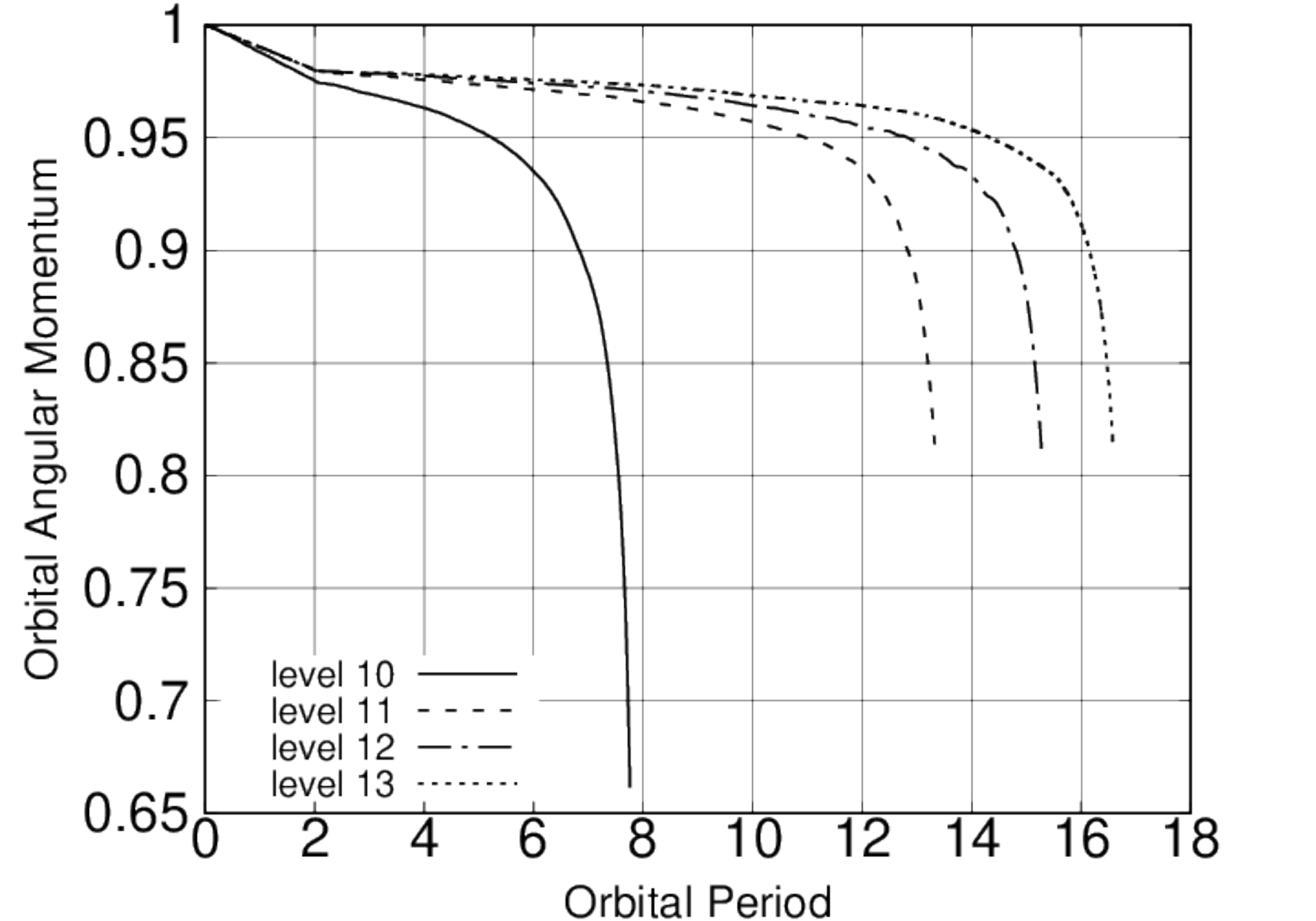}
		\caption{}
	\end{subfigure}
	\caption{The separation between the stars' centers of mass is depicted in (a), normalized to the initial separation. The donor mass loss rate is shown in (b), normalized to
one donor mass per initial orbital period. The orbital angular momentum is shown in (c), normalized to the initial orbital angular momentum. The time coordinate for all three plots is shown in units of the initial orbital period.}
	\label{fig:convergence-plots}
\end{figure*}

Figure~\ref{fig:convergence-plots} shows the evolution of three parameters of the binary's evolution. The separation is measured between the stars' centers of mass. There are epicyclic variations in the separation, while the overall trend is down. The stars continue to get closer until the donor is tidally disrupted. The donor mass loss rate is shown on a logarithmic scale, revealing how sensitive the mass transfer rate is to resolution. Lower resolution runs transfer mass from the donor at higher rates, resulting in quicker mergers. Initially, the stars are driven into closer contact by removing 1\% of the angular momentum of each cell per orbit over a period of two orbits. This is done to mimic the driving that occurs due to gravitational radiation, just on a much shorter time scale.

One prominent effect of resolution is the time at which merger occurs. Because the time-scale for each run is different, this makes grid to grid comparisons difficult. We can, however,
investigate global properties as resoluton increases. As shown in Table~\ref{tab:levels}, the time of merger is one global property whose difference between successive resolutions grows smaller as resolution increases. The orbital separation, the orbital angular momentum, and the donor mass loss rates, in Figure~\ref{fig:convergence-plots}, also exhibit this behavior. 

\section{Conclusion and Outlook}
\label{sec:conclusion}
The resolution study for the production-level simulation demonstrates that global properties such as the time of merger and mass loss rate \octo
 are resolution dependent, however, their successive differences with increasing resolution grows smaller. This is an
important verification of the simulation results.
By collecting performance data during these long-running simulations and some short runs, we were further able to determine the overheads of the frameworks utilized by \octo, as well as the overhead of gathering all the data using APEX.

We have been able to show that the overhead of collecting performance data did affect the overall runtime performance of \octo only slightly:
The difference in performance between runs with and without APEX was only around $1\%$.
The measured AGAS overhead during the short runs suggests that on more than $128$ compute nodes, less than $0.025\%$ of the total runtime is spent executing the AGAS code.
During the resolution study, the measured AGAS overhead was around $0.17\%$.
The discrepancy here is probably due to a different number of compute nodes and due to the increased scenario size and the dynamic behavior, refining the grid data structures with evolving simulation time.

The performance of the HPX AGAS itself and the low runtime overhead of measuring overheads with HPX and APEX underpin the viability of using the asynchronous task-based framework HPX for real-world applications. For the first time, we have obtained convincing numbers for large, dynamic production-scale runs and scenarios. This extends the proof of applicability of our approach from large but short-running and thus static runs to large long-running dynamic simulations with frequent re-gridding and adaptive mesh refinement.

Furthermore, we have investigated the energy consumption of running \octo.
This will provide us with an extra metric for further optimization of \octo. For large-scale simulations we have the responsibility to not only consider scaling and wall-clock time but to do science with less economical and environmental impact.
\medskip

The I/O demands of \octo at higher levels of refinement dominates the execution time. 

\section{ACKNOWLEDGMENT}

This research used resources of the National Energy Research Scientific Computing Center, a U.S. Department of Energy Office of Science User Facility operated under Contract No. DE-AC02-05CH11231; Louisiana Optical Network Infrastructure; and by Lilly Endowment, Inc., through its support for the Indiana University Pervasive Technology Institute. This work was supported by NSF Award 1814967 and the FA8075-14-D-0002/0007 grant.

\subsection{Supplementary materials}
Build scripts\footnote{\url{https://github.com/diehlpk/PowerTiger}}; input files\footnote{\url{https://doi.org/10.5281/zenodo.3751820}} and software version; and slurm files\footnote{\url{https://doi.org/10.5281/zenodo.3753539}} are available to reproduce the results.

\section*{Copyright notice}
\textcopyright 2021 IEEE. Personal use of this material is permitted. Permission from IEEE must be obtained for all other uses, in any current or future media, including reprinting/republishing this material for advertising or promotional purposes, creating new collective works, for resale or redistribution to servers or lists, or reuse of any copyrighted component of this work in other works.

\footnotesize
\bibliographystyle{unsrt}  
\bibliography{references,hpx}


\end{document}